\documentclass[journal=nalefd, layout=twocolumn, email=false]{achemso}
\setkeys{acs}{articletitle = true}

\usepackage[T1]{fontenc}
\usepackage[utf8]{inputenc}

\usepackage{bm,amssymb}
\usepackage{graphicx}

\usepackage{amsmath}
\usepackage{upgreek}
\usepackage{braket}

\title{Quantitative nanoscale MRI with a wide field of view}
\author{F. Ziem}
\affiliation{3rd Physical Institute, University of Stuttgart, Germany}
\alsoaffiliation{Institute for Quantum Science and Technology, IQst}
\alsoaffiliation{Centre for Applied Quantum Technologies}
\author{M. Garsi}
\affiliation{3rd Physical Institute, University of Stuttgart, Germany}
\alsoaffiliation{Institute for Quantum Science and Technology, IQst}
\alsoaffiliation{Centre for Applied Quantum Technologies}
\author{H. Fedder}
\affiliation{3rd Physical Institute, University of Stuttgart, Germany}
\alsoaffiliation{Institute for Quantum Science and Technology, IQst}
\alsoaffiliation{Centre for Applied Quantum Technologies}
\alsoaffiliation{Swabian Instruments GmbH, Frankenstr. 39,\\71701 Schwieberdingen, Germany}
\author{J. Wrachtrup}
\affiliation{3rd Physical Institute, University of Stuttgart, Germany}
\alsoaffiliation{Institute for Quantum Science and Technology, IQst}
\alsoaffiliation{Centre for Applied Quantum Technologies}

\begin{document}

\maketitle

\section{Abstract}

Magnetic spin resonance is a key non-invasive sensing and imaging technique across the \mbox{life-,} material- and fundamental sciences with further medical and commercial applications.
Recent advances using paramagnetic color centers enable magnetic resonance down on the nanoscale, and sensitivity to single molecules is in sight. Ensemble sensing and wide field imaging improve sensitivity and acquisition speed, but may suffer from inhomogeneous spin control fields, produced by e.g. microstructures in integrated devices and limited spatial resolution.
Here we demonstrate multiplexed nuclear magnetic resonance imaging using diamond nitrogen-vacancy centers in such adverse conditions. We image thin films of calcium fluoride down to $1.2\,\mathrm{nm}$ in thickness with a spatial resolution of $250\,\mathrm{nm}$. This corresponds to a net moment of about 140 nuclear spins within the sensing radius of a given NV in the ensemble. 

\section{Introduction}

The nitrogen-vacancy (NV) color center in diamond is a paramagnetic defect, which provides access to magnetic resonance experiments with single spins. In the negatively charged state, its photophysics result in different fluorescence intensities, conditional on the electronic spin state, which enables optically detected magnetic resonance using standard microscopes. In addition, the spin state is polarized during laser illumination of the defect.
Direct control over the NV spin is exerted by microwave (MW) irradiation, and various experimental schemes have been employed to use individual or ensemble NVs to probe magnetic\cite{Casola2018} and electric fields\cite{Dolde2011}, as well as pressure \cite{Momenzadeh2016} and temperature \cite{Neumann2013}. Remarkably, these capabilities persist over a wide range of conditions, including ambient conditions and e.g. cellular environments, providing magnetic sensing and imaging capabilities in lab-on-a-chip devices \cite{Steinert2013, Glenn2015}.
Using NV centers, nuclear magnetic resonance (NMR) spectroscopy has been demonstrated
on less than 100 nuclei\cite{Staudacher2013, Mamin2013, Lovchinsky2017} and recent improvements of the underlying
measurement schemes enable sub-part-per-million resolution in the acquired spectra \cite{Bucher2017, Pfender2017}. 
The detection of these NMR signals is predominantly based on dynamical decoupling (DD) sequences, which consist of a train of $N$ $\pi$-pulses applied to NVs in a coherent superposition of their spin states (Fig \ref{fig:setup}b). Similar to a rotating frame, these sequences put the NVs in a toggling frame in which the NVs experience a magnetic field with a net $z$-component when the spacing $\tau$ between consecutive pulses is closely matched to the nuclear Larmor frequency, \( \tau \approx (2 f_\mathrm{nuc})^{-1} \). The $z$-component induces a phase in the NV superposition which can be mapped into spin state populations and read out optically. The phase of these lock-in sequences with respect to that of the NMR signal is usually random, which results in decoherence of the NV sensors, associated with a dip in the optical contrast $C$ upon readout . When the nuclear spins are homogeneously distributed above the diamond surface, the mean square magnetic field amplitude at the position of an NV at depth $d$ below the surface $B_\mathrm{rms}^2$ is proportional to $d^{-3}$. The relative NMR contrast of the spectral feature is described by \cite{Pham2016}
\begin{equation} \label{eq:remaining_contrast}
	C \approx \exp\mathopen{}\left( -\frac{2}{\pi} \, B_\mathrm{rms}^2(d) \, K(N\tau) \right) \,,
\end{equation}
where the line shape $K(N\tau)$ is a convolution of the filter created by the DD sequence and the spectral density of the nuclear magnetic signal.
Scanning probe experiments using single NV centers have previously imaged nuclear spin densities on the nanoscale \cite{Rugar2014a, Haberle2015}, with an inherent drawback of slow image acquisition speeds due to long interrogation times at each pixel.
This limitation can be overcome in a wide field imaging mode to record the fluorescent response of a large ensemble
of NV centers in parallel \cite{Steinert2013, Glenn2015, DeVience2015, Simpson2017}.

To unleash the full power of parallelized sensing, it is key to reliably control the ensemble of NVs over its entire volume. The large number of pulses used in lock-in detection requires robust spin manipulation which is immune to inhomogeneities of the MW drive strength and magnetic resonance frequencies. Inhomogeneities are inherent to the wide field imaging mode and ensembles in general, and are caused by i) the microwave structures, especially in tightly integrated devices, and ii) the inhomogeneity of the static magnetic field, hyperfine coupling and orientations of the spin quantization axis. 
Here we focus primarily on the first point and demonstrate how homogeneous control over spins across a gradient of the control field  can be achieved by pulse shaping. 
The problem of inhomogeneous MW fields has also been addressed using different types of microwave structures, aiming to manipulate single and ensemble spins on length scales up to millimeters \cite{Sasaki2016, Eisenach2018}. The most homogeneous control has been obtained through ring-shaped resonators with diameters larger than the target area, but usually result in relatively low strength of the driving field. In addition, these structures can hardly address NV centers oriented perpendicular to the diamond surface in samples with (111) orientation. 
An extreme case of inhomogeneous control fields has been encountered with MW irradiation through the write pole of a hard disk head \cite{Jakobi2016a}, but at the same time extraordinarily high Rabi frequencies were achieved. In this setting, it is highly desirable to be able to enlarge and shape the volume in which deterministic spin control is available.

In a setting with exaggerated MW inhomogeneities, we first illustrate the problem of inhomogeneous control and then ensure that specifically tailored optimal control pulses establish homogeneous control. We apply these pulses to perform nuclear magnetic resonance imaging (MRI) of fluorine nuclei in patterned calcium fluoride thin films with a thickness down to  $1.2\,\mathrm{nm}$ over an area of $60 \times 80\,\upmu\mathrm{m}^2$. At the same time, we use the gradient of the microwave across the field of view to improve the spatial resolution of our imager by a factor of 1.3 to $250\,\mathrm{nm}$.

\section{Experimental Setting}

\begin{figure}
\centering
\includegraphics{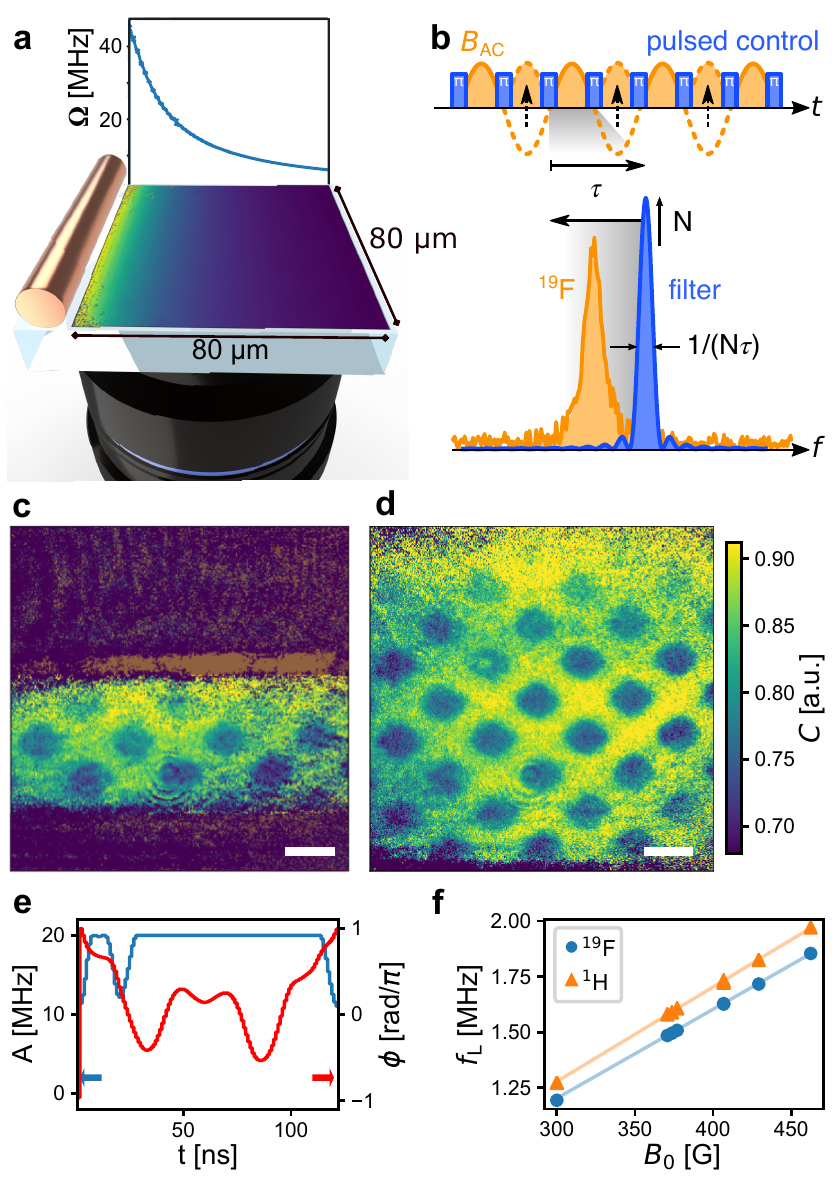}
\caption{\textbf{Experimental setting.} \textbf{a}, A $20\,\upmu\mathrm{m}$ thick wire is placed next to the field of view for the irradiation of microwaves. The $x^{-1}$ decay of the field amplitude away from the wire results in a distribution of NV Rabi frequencies across the field of view from which we extracted a mapping of the Rabi frequency vs. position.  \textbf{b}, Dynamical decoupling lock-in sensing utilizes a train of $\pi$ at distance $\tau$ to lock into a signal with period $T = 1/f = 2\tau$. This creates a filter which scales in width as $1/(N\tau)$ and in height as $N$. \textbf{c, d}, Lock-in nuclear magnetic resonance imaging of $^{19}$F nuclei in patterned CaF$_2$ on the diamond surface (dark blue squares in the image) using rectangular and optimal control pulses, respectively. Scale bars are $10\,\upmu\mathrm{m}$. \textbf{e}, Amplitude and phase of an optimal control $\pi_x$-pulse. \textbf{f}, Zeeman-shift of  $^1$H and of $^{19}$F Larmor frequencies. Standard errors are within symbol size. The solid lines correspond to literature values. }
\label{fig:setup}
\end{figure}

We employed a geometry with a thin wire with a diameter of $20\,\upmu\mathrm{m}$ as a test bed for gradients in the driving field encountered in tightly integrated systems or volume sensors, see Fig \ref{fig:setup}a. The ensemble in our study consisted of shallow, implanted NV centers with a mean depth smaller than $10\,\mathrm{nm}$ below the diamond surface. Their fluorescent response to the applied measurement sequences was imaged across a field of view of $80 \times 80 \upmu\mathrm{m}^2$ using a CCD camera. We first mapped the NV Rabi frequencies to position and registered the strong dependence of the MW amplitude $B_1$ in  $y$-direction perpendicular to the wire, but little variance in the $x$-direction parallel to the wire. In this way, we were able to perform spatially resolved imaging of the NV responses or to average the results parallel to the wire to produce one-dimensional profiles when required, e.g. to obtain the Rabi frequency at different distances from the wire (Fig. \ref{fig:setup}a). The Rabi frequency $\Omega$ of the subset of [111] NVs in $yz$ planes at distance $y$ from the wire is given by 
\[
\Omega(y) = \sqrt{ \frac{\mu_0^2 I^2 \gamma^2}{32 \pi^2}\cdot \frac{2 + (\frac{y}{z_\mathrm{w}})^2}{z_\mathrm{w}^2(1 + (\frac{y}{z_\mathrm{w}})^2)^2} + \Delta^2,}
\]
where $I$ is the peak current in the wire, $\gamma$ is the NV gyromagnetic ratio, $z_w$ is the distance of the wire center from the NV-layer in $z$-direction and $\Delta$ is a possible detuning of the MW from the spin resonance, e.g. due to hyperfine interaction. 
We then applied XY16 dynamical decoupling sequences, which serve the dual purpose of suppressing decoherence due to magnetic field noise in the NV environment, but also increase sensitivity to field fluctuations within a narrowed frequency window, as shown in Fig. \ref{fig:setup}b. These sequences can be applied to detect  statistical polarization in mesoscopic spin ensembles \cite{Meriles2010} which create an oscillating magnetic field as a result of their Larmor precession. 

When using standard rectangular MW pulses, the phase cycling in XY16 enables dynamical decoupling and sensing within an error of the Rabi frequency up to about $\pm20\,\%$ at around 200 pulses, as shown below. In this way, we were able to detect the Larmor precession of $^{19}$F in CaF$_2$ patches deposited on the diamond surface, Fig. \ref{fig:setup}c, but were limited to the range imposed by the MW inhomogeneity. In order to increase the range of control, there are two general approaches. One is to further improve the phase cycles between pulses, e.g. in concatenated sequences \cite{Farfurnik2015}. These can be employed in a straight-forward manner, but the number of pulses quickly grows beyond what is necessary (and sensible) for NMR detection. A second approach is to optimize the individual pulses themselves, which can be achieved in composite pulses \cite{Levitt1986} or by using optimal control theory \cite{Borneman2010, Nobauer2014, Rose2017}. In essence, the aim of optimal control (OC) pulse shaping is to design amplitude and phase modulation of the applied control field to realize a desired operation given a range of experimental parameters and constraints. By explicitly optimizing for given experimental conditions, the pulse duration can be kept as short as possible
We employed the GRAPE algorithm \cite{Khaneja2005, Borneman2010} to engineer $\pi$ and $\pi/2$ rotations on all members of the ensemble homogeneously to replace the rectangular pulses in XY16 (see Methods section). The target ensemble comprised NV centers undergoing Rabi nutations in a range of $\beta = \Omega/\Omega_0 = 0.5 \dots 1.5$ with a reference Rabi frequency of initially $\Omega_0 = 20\,\mathrm{MHz}$. In other words, the OC pulse approximated the same effective unitary operation for spins with Rabi frequencies within a factor of 3, between $10\,\mathrm{MHz}$ and $30\,\mathrm{MHz}$. We arrived at the $\pi$-pulse shown in Fig. \ref{fig:setup}e, with a duration of $122\,\mathrm{ns}$ and an average fidelity of 0.991. We optimized an adequate $\pi/2$ pulse over a slightly larger range of $\beta = 0.4 \dots 1.6$ with a duration of $198\,\mathrm{ns}$ and an average fidelity of 0.992. Anticipating a more detailed characterization of the OC pulses below, Figure \ref{fig:setup}d demonstrates that they were able to extend the range of the NMR detection to almost the complete field of view. Measurements across a range of static magnetic field strengths confirmed the Zeeman-shift of the fluorine signal and that of $^{1}$H in water adsorbed on the diamond surface, as shown in Fig \ref{fig:setup}f.

\section{Results and Discussion}

\begin{figure}
	\centering
	\includegraphics{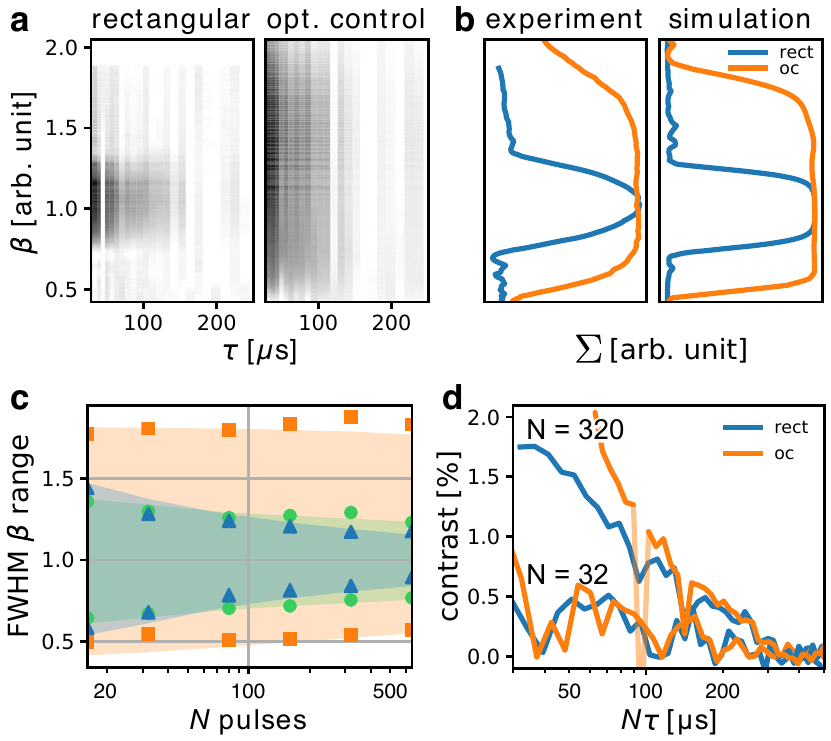}
	\caption{\textbf{Decoupling performance.} We performed XY16-$N$ with $N$ pulses of either rectangular, knill or optimal control type. a) Coherence decay data averaged parallel to the wire (XY16-320). The remaining amplitude for each $\tau$ is intensity coded in the images (darker for higher coherence). Optimal Control pulses decouple over a notably increased range of the MW gradient. b) Numerical integral $\Sigma$ over experimental and simulated data. The OC pulses decouple almost homogeneously across their designed bandwidth. c) Scaling of the effective decoupling bandwidth in XY16 using three different types of pulses: rectangular (blue triangles), OC (orange squares) and Knill pulses (green circles). The shaded areas were obtained from numerical simulations. d) Experimental $T_2$ decay curves in decoupling experiments. Matching shapes were observed with differences caused by the different durations of the pulses, including a pronounced dip for $N = 320$/OC due to hyperfine interaction with the NV nitrogen nucleus. For $N = 32$, the $^{13}$C spin bath leads to pronounced NMR dips.}
	\label{fig:DD_scaling}
\end{figure}

To ensure that the OC pulses resulted in the same decoupling behavior across their whole design range,
we compared them to rectangular pulses in XY16 measurements. We applied XY16-$N$ with $N$ pulses, ranging from 16 to 640. For each $N$, we varied the interpulse distance up to a total of $N\tau = 500\,\upmu\mathrm{s}$ free evolution time (excluding the respective pulse durations). In all cases, we used the same OC $\pi/2$ pulses to initialize the coherent superposition between NV eigenstates and for the final projection. We averaged the recorded data along the axis parallel to the wire and summarize the resulting decay curves for $N =320$ in Fig. \ref{fig:DD_scaling}a. Example curves averaged over the pixels in the range $\beta = 0.9 \dots 1.1$ are shown in Fig. \ref{fig:DD_scaling}d.
As anticipated, the rectangular pulses performed well around $\Omega_0 = 20\,\mathrm{MHz}$, but once the Rabi frequency notably differed from that value, pulse errors dominated the sequence and coherence was quickly lost. Since the curves in these two extreme conditions are qualitatively different, it is not straight forward to characterize them with a common model function. To compare the effectiveness of the decoupling we numerically integrated over the data points to obtain the enclosed area $\Sigma$, which we took as a measure of the quality of the decoupling. Since the integral $\Sigma = \int_0^\infty A\exp(-\tau/T)\mathrm{d}\tau = AT$ for a monoexponential decay, it can be used for robust comparison (Supplement for details).
In Fig. \ref{fig:DD_scaling}b, we show the $\Sigma$ profiles corresponding to the decays in Fig. \ref{fig:DD_scaling}a and and numerical simulations of the sequence, respectively. In the simulations, we only included pulse errors and $^{15}$N hyperfine interaction. From the good agreement between experiment and simulation, we conclude that the range of effective decoupling is dominated by pulse errors and not by relaxation due to the spin bath or additional hyperfine interaction. While we have focused on rectangular and OC pulses, we also included a composite Knill pulse, which consists of five consecutive $\pi$-pulses at different phases\cite{Ryan2010}. With a total duration of $125\,\mathrm{ns}$, it is comparable to the duration of the OC pulse, but is effective only across a much narrower range of $\beta$. Compared to rectangular pulses, Knill pulses were effective over a slightly larger range at relevant pulse numbers and exhibited a more flat-top shape of the $\Sigma$ profiles. In Fig. \ref{fig:DD_scaling}c we compare the full width at half max of the $\Sigma$ profiles for all measurements and corresponding simulations to determine how the bandwidth of the error correction scales for all three types of pulses. The range of Rabi frequencies, over which the rectangular decoupled well, narrowed down with increasing number of pulses and reaches a value of $\Omega_0\pm15\,\%$ for 640 pulses. The optimal control pulses performed equally well across all applied pulse numbers. 

\begin{figure*}
		\includegraphics{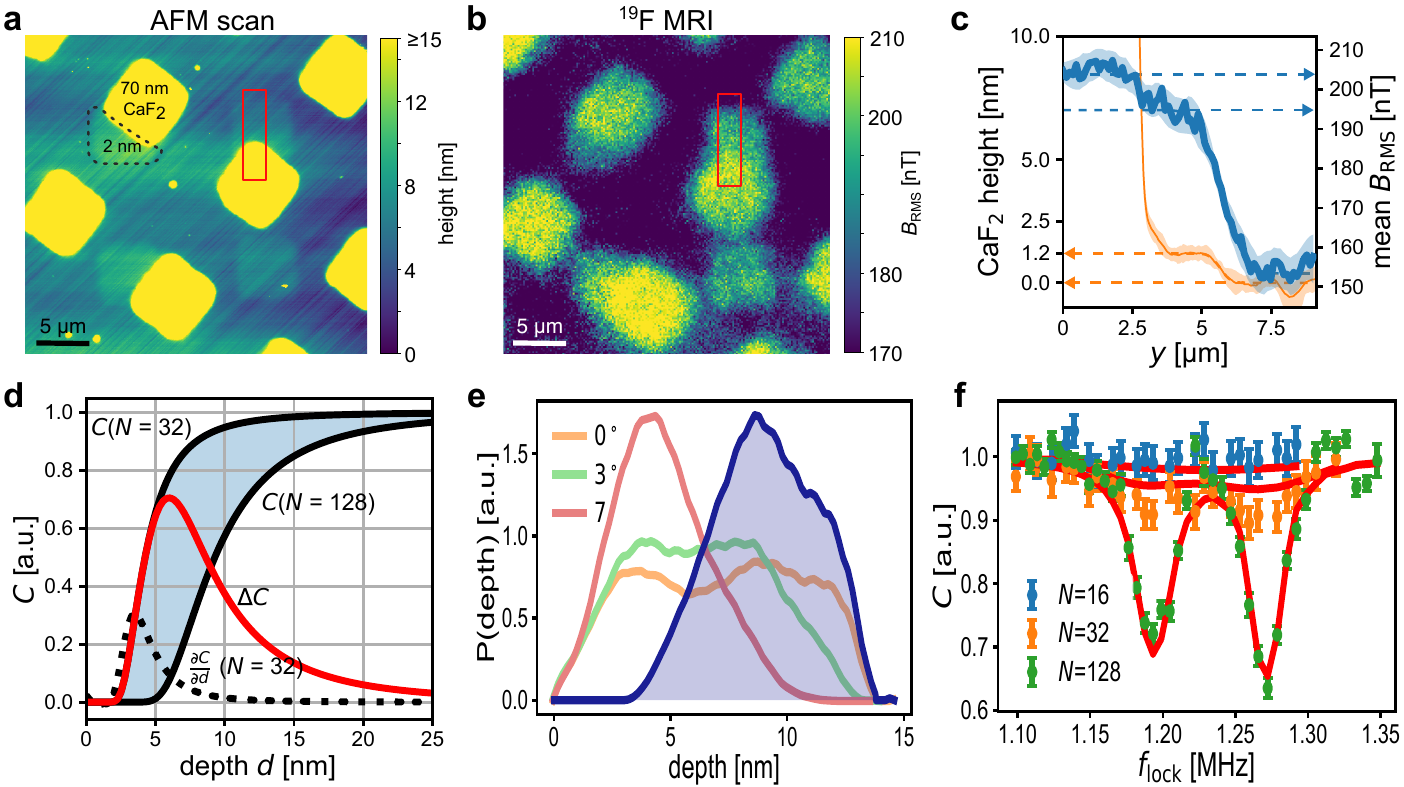}
	\caption{\textbf{Nuclear magnetic resonance imaging.} The diamond surface was patterned with CaF$_2$ islands. a) AFM scan of the diamond surface. b) Approximate effective magnetic field amplitude $B_\mathrm{rms}$ measured with XY16-256. c) Profiles of the surface and  $B_\mathrm{rms}$ along a transition from $70\,\mathrm{nm}$ CaF$_2$ to $1.2\,\mathrm{nm}$ CaF$_2$ to the bare diamond. d) Calculated on-resonance lock-in contrast to reconstruct the distribution of NV depths in the ensemble. The coherence loss due to the nuclear phase noise depends on NV depth and the duration of the lock-in (given by the number of DD pulses). Single measurements have a dynamic range illustrated by the dotted line, while combined measurements cover intermediate depths using CTRIM simulations as a basis. e) Recovered distributions of NV depths with mode at $8.6\,\mathrm{nm}$ and standard deviation of $2.2\,\mathrm{nm}$. f) Ensemble averaged data for NVs below $70\,\mathrm{nm}$ CaF$_2$ and fit using the recovered distribution of depths.
		}
	\label{fig:mri}
\end{figure*}

Having established that the optimal control pulses can replace rectangular pulses in dynamical decoupling sequences, we apply them in NMR detection. In this instance, we deposited two patterned grids of CaF$_2$ thin films on the diamond surface using electron-beam physical vapor deposition (EBPVD) of $1.20(5)\,\mathrm{nm}$ and $70\,\mathrm{nm}$ thickness, as confirmed by atomic force microscope (AFM) measurements, Fig. \ref{fig:mri}a. We detected the fluorine nuclei within these thin films, Fig. \ref{fig:mri}b, as well as the proton signal of adsorbed water on the surface. The results shown here also include a constant $^{19}$F signal across the whole diamond, which remained after initial reference measurements with a polytetrafluoroethylene (PTFE) coating, but was absent before. Notably, the thin deposited layers are clearly visible against this background. 
At first, we empirically fitted the measured spectral features including a finite $T_2^*$, as detailed by Pham \textit{et al.}\cite{Pham2016}, resulting in apparent $B_\mathrm{rms}$ values given in  Fig. \ref{fig:mri}b,c. While these results are of the right order of magnitude, within an ensemble, the average response of the NV sensors strongly depends on NV depths below the diamond surface. Since the depth distribution due the implantation at $2.5\,\mathrm{keV}$ is not known exactly, we estimate it from the NMR measurements. Within the assumptions of the DD NMR model \cite{Loretz2014, Pham2016}, by choosing a single appropriate number of decoupling pulses $N$ the depth of a single NV can be determined with Angstrom accuracy. The sensitivity of the NMR contrast to different NV depths relates to the derivative $\partial C/\partial d$, illustrated by the dotted line in figure \ref{fig:mri}d. The NMR contrast observed in an ensemble measurement with $N$ pulses is an average over the contrast of NVs distributed across a range of depths:
\begin{equation} \label{eq:avg_contrast}
	c^\prime_N = \int_0^\infty P_d(d) C_N(d)\,\mathrm{d}d\,,
\end{equation}
where we emphasize the role of $N$ by adding it as a subscript to the contrast $C$. Due to the non-linear relation between NV depth and NMR contrast, the NV density can be reconstructed by applying NMR measurements with a range of number of pulses $N$. For two different $N = A, B$, the change in the observed contrast is given by
\begin{equation} \label{eq:contrast_diff}
	\Delta c^\prime = \int_0^\infty P_d(d)[C_A(d) - C_B(d)]\,\mathrm{d}d\,,
\end{equation}
as illustrated in figure \ref{fig:mri}d. Pairwise combinations of measurements define a set of equations from which the distribution of NV depths $P_d$  can be inferred by fitting a model distribution or in a parameter-free approach. 

To reconstruct the distribution of NV depths of the $2.5\,\mathrm{keV}$ implantation used in our experiment, we combined measurement data with $N$ = 16,  32, 128, 160, and 256. For each of these measurements, we separately averaged the NMR signal from NVs below $70\,\mathrm{nm}$ CaF$_2$, $1.2\,\mathrm{nm}$ CaF$_2$ and without CaF$_2$, each on an area of about $2\,\upmu\mathrm{m}^2$. The nuclear spin density at the surface was composed of i) the protons in adsorbed water, ii) the fluorine nuclei in the residue of PTFE, and iii) the fluorine nuclei where CaF$_2$ had been deposited. The effective square magnetic field $B_\mathrm{rms}^2$ is a sum over the contributions of the individual nuclei, which allowed us to separately model each of these contributions as a layer of finite thickness. Knowing the measured thicknesses of the CaF$_2$ patches and the the bulk density of fluorine nuclei in CaF$_2$ of about $4.9\times10^{28}\,\mathrm{m}^{-3}$, we left the spin densities and thicknesses of the proton and residual fluorine layers as free parameters. The distribution of NV depths has been found to be well described by CTRIM simulations of the nitrogen implantation \cite{FavarodeOliveira2015}, where the distribution of depths was obtained by step-wise etching of the diamond followed by confocal NV counting. CTRIM takes the structure of the diamond lattice into account and includes channeling of ions in the implantation process, which may lead to bimodal distributions of the NV depths. This effect depends on the angle of the ion beam against the surface normal and we took simulations for $0^\circ$, $3^\circ$ and $7^\circ$ as the basis for the reconstruction. In addition, we added a $\cos^2$-shaped depletion zone at the surface, to account for instability of charge state and NV formation. The relative weights of the three CTRIM profiles, as well as the onset and width of the depletion zone were left as free parameters. We combined the ensemble average measurement data, the spin layer model and the model depth distribution in one least square optimization and recovered the depth distribution shown in figure \ref{fig:mri}e. Example data with $N = 16, 32, \text{ and } 128,$ as well as fits for the $70\,\mathrm{nm}$ thick layer are shown in \ref{fig:mri}f. The fit did not fully reproduce the dip for $N = 32$, which shows that the reconstructed density is missing probability mass closer to the surface, in accordance with single NV results, for which NV depths below $2\,\mathrm{nm}$ have been reported \cite{Loretz2014}. On the other hand, the mode of the distribution around $8.6\,\mathrm{nm}$ and standard deviation of $2.2\,\mathrm{nm}$ is in accordance with statistics of NV depths of the same implantation energy obtained from NMR measurements using single NVs \cite{Pham2016}. 
We obtained thicknesses and spin densities of $1.5(4)\,\mathrm{nm}$ and $6.1 \times 10^{28}\,\mathrm{m}^{-3}$ for the proton layer, and $4.0(2)\,\mathrm{nm}$ and $1.7\times10^{28}\,\mathrm{m}^{-3}$ for the residual PTFE, respectively.
Due to e.g. the dependence of the depth distribution on implantation angle, we expect that characterization of each ensemble NV sensor is necessary before quantitative measurements on near-surface samples is feasible. Once a probability distribution of NV depths $P_d(d)$ has been obtained, the distribution of NMR contrast $P_C(C)$ results from a change of variables within equation \ref{eq:remaining_contrast}. We rewrite eq. \ref{eq:remaining_contrast} as $C = \exp(-M d^{-3})$, with inverse $d(C)=\sqrt[3]{-M/\ln(C)}$, where $M$ includes all influences but NV depth, and obtain the distributions of NMR contrasts
\begin{equation}
	P_C(C) = \frac{P_d(d(C)) M}{3C(-M/\ln(C))^{2/3}\ln(C)^2},
\end{equation}
which evaluates in the open inverval
\[ C \in (\lim\limits_{d \to 0} C = 0, \lim\limits_{d \to \infty} C = 1). \]
The response measured at each pixel is the average over the responses from individual NV centers which are located within the radius of the optical point spread function around the pixel. 

\begin{figure}
	\centering
	\includegraphics{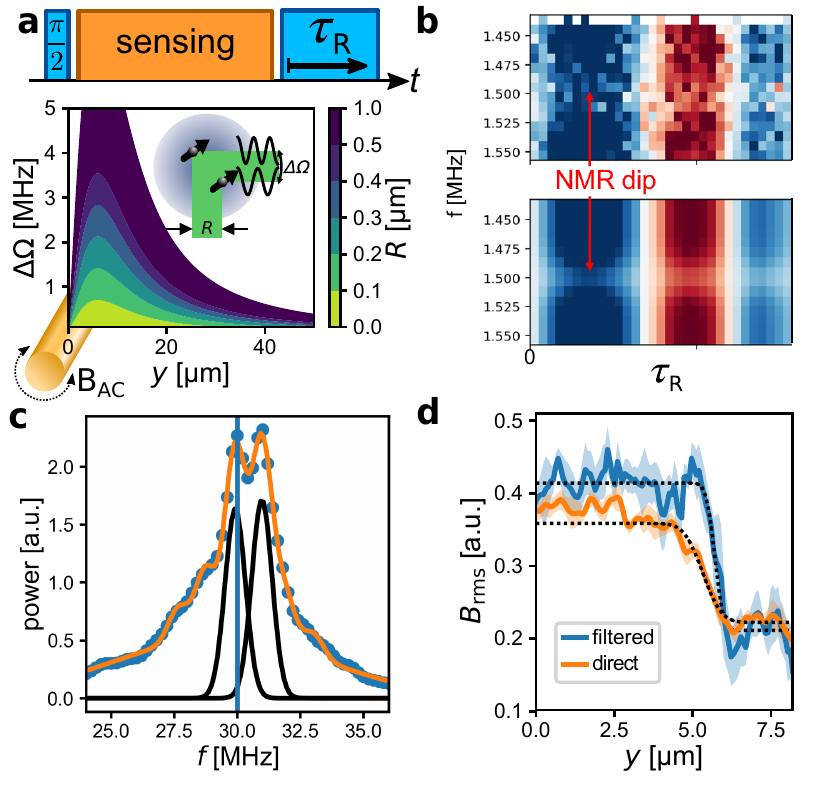}
	\caption{\textbf{Superresolution in MW gradient.} a) A spin state projection pulse with variable duration $\tau_\mathrm{R}$ results in an oscillating sensor response, modulated by the sensing result. Discrimination between spectral components with resolution $\Delta \Omega$ translates to spatial resolution $R$ at a given distance $y$ from the wire.  b) The Rabi readout signal is modulated in amplitude by the result of an $^{19}$F NMR signal at $1.5\,\mathrm{MHz}$. Data (top) and direct fit (bottom) from a single pixel. c) At each pixel, a spectrum of the Rabi oscillations from surrounding NVs enables band filtering to the signal originating from a narrow spatial range perpendicular to the wire. Within sparse areas of the ensemble, individual components can be identified. d) Fitting the NMR signal after filtering the spectra recovers sharpened steps at CaF$_2$ edges.
		}
	\label{fig:gradient_application}
\end{figure}

In a final step, we turn to utilizing the gradient within the control field. While we introduced the inhomogeneous driving field to demonstrate its detrimental effect on ensemble sensing and showed how the gradient may be compensated, we can also take advantage of having access to a continuous range of Rabi frequencies. In the present case, we can use the gradient within the Rabi frequencies to improve the spatial resolution perpendicular to the gradient \cite{Shin2010, Fedder2011a}, see Fig. \ref{fig:gradient_application}a. We modify the spin projection and replace the $\pi/2$ pulse with a pulse of variable duration $\tau_\mathrm{R}$, effectively appending a Rabi measurement to the sensing sequence. As a result, the amplitude of the observed Rabi frequencies is modulated by the sensing result and we refer to this technique as Rabi readout (RR). By extracting the spectra of the Rabi oscillations and subsequent band-filtering to a narrow range around the expected Rabi frequency in the gradient, the components of different NVs positioned along the gradient can be separated from each other. Given sufficient spectral resolution $\Delta \Omega,$ NVs with a separation $R$ smaller than the diffraction limit may be spatially resolved in this way, cf. Fig. \ref{fig:gradient_application}a. The wide-bandwidth optimal control pulses are still key to being able to prepare and manipulate the ensemble of NVs across the gradient, when no additional homogeneous control field is available. We applied the RR with $\tau_\mathrm{R}$ up to $500\,\upmu\mathrm{s}$ at the end of XY16-256 NMR sequences. Fig. \ref{fig:gradient_application}b shows the initial oscillations from a single pixel and corresponding fit with a damped sinusoidal and generic Gaussian dip to model the NMR signal. We sampled $\tau_\mathrm{R}$ at 100 random values to take advantage of sparse sampling \cite{Arai2015}. For unevenly sampled data, the Lomb-Scargle spectrum can be used as an alternative to Fast-Fourier Transforms\cite{VanderPlas2017} and is equivalent to $\chi^2$ weighted spectra of least-square sinusoidal fits at preselected probe frequencies. The spectrum can be normalized by the model $\chi^2$, which introduces a non-linear enhancement when the data closely match a probe frequency (Supplement for details). With the known MW gradient, components at different frequencies in the spectra can be attributed to a position within the gradient. An example spectrum is shown in Fig. \ref{fig:gradient_application}c, where the gradient was close to $1\,\mathrm{MHz}/\upmu\mathrm{m}$ at $20\,\upmu\mathrm{m}$ from the wire. In some areas, we observed reproducible, distinct peaks within the spectra, which we tentatively attribute to individual NV centers within the relatively sparse ensemble. In principle, this allows localization of NVs by decomposition of the spectra into individual components. The uncertainty of $0.05\,\mathrm{MHz}$ in the line centers in Fig. \ref{fig:gradient_application}c  corresponds to about $50\,\mathrm{nm}$ precision in spatial resolution. As a general procedure for dense ensembles, we continued with the following analysis. We numerically filtered the obtained spectra around the expected Rabi frequencies within a narrow band given by the gradient and pixel size and in this way limited the signal at each pixel to the components mainly contributed by nearby NVs. E.g. within the above example, our physical pixel size of $160\,\mathrm{nm}$ resulted in a filter width of $0.16\,\mathrm{MHz}$, from which we expect a spatial resolution of about $0.2\,\upmu\mathrm{m}$, cf. Fig. \ref{fig:gradient_application}a. The spectral filtering results in lock-in spectra at each pixel, which we fit with a Gaussian dip. Fig. \ref{fig:gradient_application}d shows a comparison of step widths obtained by direct fitting and after filtering. 
We compared the step widths at six different CaF$_2$ patches and found an average step width of $\sigma_\mathrm{direct} = 0.45(9)\,\upmu\mathrm{m}$ without filtering and $\sigma_\mathrm{filtered} = 0.26(8)\,\upmu\mathrm{m}$ after filtering. This compares to an expected effective step width of $0.35\,\upmu\mathrm{m}$. This latter value is composed of the width of the step in AFM measurements, $0.30(3)\,\upmu\mathrm{m}$, the non-linear response of the NMR contrast $C$ to the thickness of a finite spin layer, and finally convoluted with $\sigma_\mathrm{opt} \approx 0.165\,\upmu\mathrm{m}$ due to optical diffraction. The non-linear amplification using the $\chi^2$ weighting in the spectra increased the step height within the filtered results (Fig. \ref{fig:gradient_application}d) and may also result in sharpened edge detection. In summary, after spectral filtering a step width compatible with the AFM result is recovered, about a factor 1.3 below the diffraction limit. 

\section{Conclusions}

We have shown in this work that bolstering dynamical decoupling sequences with robust pulses enables control of spin ensembles even in adverse experimental conditions in terms of driving field inhomogeneity. Composite pulses may already improve the performance of decoupling across an increased range of driving strengths. In our case, Knill pulses in XY16 sequences led to uniform control over a range of $\Omega_0\pm 20\;\%$. In case of a well characterized system with known Hamiltonian, optimal control pulses can further increase the bandwidth at the same duration as a composite pulse, which may give a decisive advantage in applications which set tight boundaries on duration or power of the applied pulses. Although we did not focus on this second aspect of low energy consumption, the optimization process can be modified to favor pulses with low amplitudes. Since power scales quadratically with the amplitude of the control field, optimal control pulses could be especially useful in low temperature experiments were little dissipation of heat into the cryostat is a key point to consider. In other applications, such as double resonance, putting the focus on pulses with large frequency bandwidths will enable to manipulate a target spin which changes its resonance frequency frequently, e.g. due to strong hyperfine coupling or ensembles of such spins. Our chosen geometry, using a wide field approach, enabled us to characterize the performance of an optimal control pulse quickly. The gradient in MW amplitude could in the future be accompanied by a second gradient at right angle, e.g. of the static magnetic field, to map the performance of the pulses across a range of detunings in parallel. This hints at the potential application of wide field imagers for multiplexed probing of any quantity accessible to the NV (magnetic and electric fields, temperature, strain) across a large set of combinations of at least two controlled parameter gradients, like DC and AC field strengths and density as well as chemical composition of a target environment.
We applied the optimized pulses in NMR microscopy and imaged thin films containing fluorine resulting in an effective signal of about 140 nuclei. From the experimental signal-to-noise ratio, we expect to be able to clearly image layers down to $0.5\,\mathrm{nm}$ in thickness, corresponding to about 80 nuclei, within four hours measurement time.
We have also introduced a modified readout scheme, in which the usual final projection onto the NV eigenstates using a $\pi/2$-pulse is replaced by Rabi-type driving and readout. The result of the measurement sequence can be encoded into the amplitude and/or phase of the periodic NV response. As shown, this scheme may be used to increase the spatial resolution, when a known mapping between Rabi frequency and position exists. Improved spatial resolution is achievable within modest gradients, but improvements in resolution are expected closer to the wire, where the gradient is steeper, or by increasing the MW power, which we operated considerably below saturation. Numerical simulations of Rabi readout NMR measurements with a gradient of $5\,\mathrm{MHz}/\upmu\mathrm{m}$ resulted in a resolution of $40\,\mathrm{nm}$ (Supplementary Information). The Rabi readout scheme may be incorporated into existing wide field superresolution techniques or as an ad-hoc way to utilize inhomogeneities which are already present in the system. It may become a crucial tool in light of recent results, in which static and AC gradients have been applied to site-selective addressing of NV centers on the nanoscale \cite{Zhang2016a, Bodenstedt2018}. With the proposed scheme, a measurement can be performed on all sensors in parallel and spatial encoding is only added upon readout, either by Rabi driving in the MW gradient as shown here, or with a Ramsey-type modulation in a DC gradient just before the final $\pi/2$ projection pulse.

\section{Materials and Methods}

The experimental realization of the optimal control pulses was achieved by IQ modulation of
a MW source (Rhode\&Schwarz SMBV100A) using an arbitrary waveform generator (AWG) (Tektronix AWG520). 
All pulses, including the standard rectangular pulses, were generated using this combination. The
AWG was set to a clocking rate of one gigasample per second, the same time step
used in the optimal control pulses.
In parallel to pulse shaping, the AWG also controled a MW switch (MiniCircuits) between MW 
source and amplifier to suppress pulse transients and spurious signals during free evolution
times between the pulses. Finally, the generated MW signal was sent through a band 
pass filter (MiniCircuits), amplified (100S1G4, Amplifier Research) and applied to the wire. 
All MW components are operated well below their saturation powers. We determined the
modulation band width ($200\,\mathrm{MHz}$) and pulse shape after passing through the whole MW chain and found no
limitations.  
The diamond sample was an ultrapure element6 CVD grown monocrystal with natural abundance of $^{13}$C. We created NV centers by implantation of $^{15}$N$^+$ at a fluence of $10^{12}\;\mathrm{cm}^{-2}$ and energy of $2.5\,\mathrm{keV}$, followed by annealing at $800\,^\circ\mathrm{C}$ for $2\,\mathrm{h}$. From the implantation parameters, we estimate a lateral NV density of about $20\,\upmu\mathrm{m}^{-2}$. We measured an ensemble average $T_2 = 12(3)\upmu\mathrm{s}$ and estimated a sensitivity of $0.9(2)\,\upmu\mathrm{T}/(\sqrt{\mathrm{Hz}}\,\upmu\mathrm{m})$ within the central $600\,\upmu\mathrm{m}^2$ of the field of view from the signal-to-noise ratio in XY16-160 measurements.

The observed steps at the transition to the CaF$_2$ patches in NV and AFM measurements were well described by the convolution of a unit step with a normal distribution,
\[
y = c + \frac{a}{2} \mathrm{Erfc}\mathopen{}\left(\frac{x - x_0}{\sqrt{2}\sigma}\right)\,,
\]
where $\mathrm{Erfc}$ is the complementary error function, $c$ is a vertical offset, $a$ is the height of the step, $x_0$ is the center of the transition and $\sigma$ is the standard deviation of the normal distribution.

\subsection{Pulse engineering with GRAPE}

We applied gradient ascent pulse engineering (GRAPE) to unitary propagators of the spin state with the aim to obtain robust pulses which result in $\pi$ rotations irrespective of the initial state. In doing so, our aim was to obtain universal replacements for standard rectangular pulses, which improve the robustness of existing pulse sequences. We followed the example set by Borneman \textit{et al.} \cite{Borneman2010}, but offer a complementary view by focusing on large inhomogeneities $\beta$ in the $B_1$ driving field instead of large detunings $\Delta$ from the central resonance frequency. Similar studies, albeit with smaller range of $\beta$ have been published\cite{Kobzar2004, Nobauer2014}. Detunings may in principle be covered by increasing the MW power, but inhomogeneities in the MW field can only be covered by pulse design. 
In the secular approximation and rotating frame, the Hamiltonian for one NV in the ensemble is
\begin{equation}
	\mathcal{H} = DS_z^2 - (\gamma B_0 + \Delta )S_z  - \beta\gamma(B_x(t)S_x + B_y(t)S_y).
\label{eq:hamiltonian}
\end{equation}
The first two terms describe constant contributions, where $D$ is the axial zero field splitting parameter of the NV, $B_0$ is a constant field aligned along the NV axis and $\Delta$ is the sum of instantaneous sources of detuning from the applied MW frequency. Neglecting perpendicular terms of the hyperfine coupling, $\Delta$ includes the parallel hyperfine coupling to the intrinsic $^{15}$N nucleus of the NV and randomly distributed $^{13}$C nuclei (at $1.1\,\%$ natural abundance). It also includes possible offset of the MW frequency from the NV resonance. Microwave control over the NV spin is expressed by the last term with time-dependent $x$- and $y$-component 
The optimal control pulses are a pair of step-wise constant control amplitudes for the $x-$ and $y-$component of the MW field, equivalent to modulating its amplitude and phase. For each time step $k = 1 \dots N$, a propagator $U_k$ can be obtained via the NV Hamiltonian and the product $U_\mathrm{oc} = U_\mathrm{N} U_\mathrm{N-1} \dots U_1$ is the propagator for the complete optimal control pulse. We search for control amplitudes such that $U_\mathrm{oc}$ approximates a target propagator $U_\mathrm{t}$, e.g. that of an ideal $\pi$ pulse. Using the fidelity $\Phi,$ \[0 < \Phi = |\mathrm{Tr}(U_\mathrm{oc}U_\mathrm{t}^\dag)|^2/(\mathrm{Tr}(U_\mathrm{t}U_\mathrm{t}^\dag)\mathrm{Tr}(U_\mathrm{oc}U_\mathrm{oc}^\dag)) < 1\] as performance functional, an optimization problem is obtained.
The large parameter space given by the piecewise constant controls prevents analytical solutions to this problem, but GRAPE offers an elegant way to find local optima by providing a search direction (gradient) derived from the propagators $U_k$ and $U_\mathrm{t}$ \cite{Khaneja2005}. Finally, in order to optimize for an ensemble of detunings $\Delta$ and $B_1$ inhomogeneities $\beta$, we calculate $\Phi$ for each possible combination and obtain an ensemble performance functional $\tilde{\Phi} = \sum p(\Delta, \beta) \Phi_{\Delta, \beta}$, in which the contributions from different elements of the ensemble can be weighted via $p(\Delta, \beta)$. The search direction of the optimization is the weighted average over the gradients obtained with each $\Phi_{\Delta, \beta}$.
In phase cycling decoupling sequences, like XY8 and XY16, phase shifted $\pi$-pulses are required (e.g. $\pi_x$, $\pi_y$). From a given optimal control $\pi$-pulse, we can create pulses which are arbitrarily shifted in phase, by simply adding this phase to the $x,y$ controls.
We built the ensemble around a central Rabi frequency of $\Omega_0 = 20\,\mathrm{MHz}$, which in itself would result in a rectangular $\pi$ pulse of $25\,\mathrm{ns}$ duration on resonance. For the $\pi$ pulse, we optimized for deviations of the Rabi frequency $\Omega$ from $\Omega_0$ in a range $\beta = \Omega/\Omega_0=0.5\dots1.5$ in steps of 0.05. We aimed at as short pulses as possible, which necessarily keeps the amplitude of the control fields high, such that possible detunings due to $^{13}$C hyperfine coupling are mostly covered and we included the hyperfine splittings due to both, $^{14}$N and $^{15}$N, i.e. $\Delta = 0, \pm1.5 \,\mathrm{and}\, \pm2.2\,\mathrm{MHz}$. Inhomogeneities of the static magnetic field $B_0$ were on the order of $1\,\mathrm{MHz}$ across the whole field of view. We employed random but smooth control fields as starting guesses, limited in bandwidth to few tens of MHz, well below the 200 MHz modulation bandwidth of the hardware. The temporal resolution within the pulse was $1\,\mathrm{ns}$. We then optimized pulses for a range of durations to find a duration for which a fidelity of $\geq0.99$ could be achieved. For example, the pulse used in our decoupling measurements and shown in Figure \ref{fig:setup}e has a calculated fidelity of 0.991 at a duration of $122\,\mathrm{ns}$. We aimed to apply the OC pulses at $B_0 \approx 400\,\mathrm{G}$, and treated the NV as an effective spin 1/2 system in the calculation. Evaluating the fidelity of a finished pulse for spin 1 in the subspace spanned by $\ket{0}$ and $\ket{1}$ yielded no significant deviation from the spin 1/2 case. We optimized $\pi/2$ pulses for a broader range than the $\pi$ pulses of $\beta = 0.3\dots0.7$, in order ensure that the $\pi$ pulse limited the width of the decoupling and not the initialization and final projection. We used the resulting $\pi/2$ pulse in all measurements (including those with rectangular $\pi$ pulses) in order to unambiguously attribute the observed differences to the different $\pi$ pulses. The $\pi/2$ pulse had a fidelity of 0.992 at a duration of $198\,\mathrm{ns}$.

\section{Acknowledgements}

The authors thank Andrej Denisenko for providing CTRIM data.

\providecommand{\latin}[1]{#1}
\makeatletter
\providecommand{\doi}
  {\begingroup\let\do\@makeother\dospecials
  \catcode`\{=1 \catcode`\}=2\doi@aux}
\providecommand{\doi@aux}[1]{\endgroup\texttt{#1}}
\makeatother
\providecommand*\mcitethebibliography{\thebibliography}
\csname @ifundefined\endcsname{endmcitethebibliography}
  {\let\endmcitethebibliography\endthebibliography}{}

\end{document}